\documentclass[12pt]{article}
\usepackage{amsmath}
\usepackage{cite}
\begin{document}
\begin{flushright}
KANAZAWA-17-07\\
September, 2017
\end{flushright}
\vspace*{1cm}

\renewcommand\thefootnote{\fnsymbol{footnote}}
\begin{center} 
{\Large\bf Possible roles of 
Peccei-Quinn symmetry in an effective low energy model}
\vspace*{1cm}

{\Large Daijiro Suematsu}\footnote[1]{e-mail:
~suematsu@hep.s.kanazawa-u.ac.jp}
\vspace*{0.5cm}\\

{\it Institute for Theoretical Physics, Kanazawa University, 
Kanazawa 920-1192, Japan}
\end{center}
\vspace*{1.5cm} 

\noindent
{\Large\bf Abstract}\\
Strong $CP$ problem is known to be solved by imposing Peccei-Quinn 
(PQ) symmetry. However, domain wall problem caused by the spontaneous 
breaking of its remnant discrete subgroup could make models invalid 
in many cases. We propose a model in which the PQ charge is assigned quarks 
so as to escape this problem without introducing any extra colored fermions. 
In the low energy effective model resulting after the PQ symmetry 
breaking, both the quark mass hierarchy and the CKM mixing could be 
explained through Froggatt-Nielsen mechanism. If the model is combined 
with the lepton sector supplemented by an inert doublet scalar and 
right-handed neutrinos, the effective model reduces to the scotogenic 
neutrino mass model in which both the origin of neutrino masses and dark 
matter are closely related. The strong $CP$ problem could be related to 
the quark mass hierarchy, neutrino masses and dark matter through the PQ 
symmetry.

\newpage
\setcounter{footnote}{0} 
\renewcommand\thefootnote{\alph{footnote}}

Strong $CP$ problem is one of serious problems in the standard model (SM),
which is suggested by the experimental bound of the electric dipole 
moment of a neutron \cite{strongcp}. 
Invisible axion models based on the Peccei-Quinn (PQ) symmetry
$U(1)_{PQ}$ are known to give a simple and interesting solution to it \cite{pq}.
Since the models predict the existence of a pseudo-scalar called axion 
\cite{axion}, which has very small mass and extremely weak 
interaction with matter, the scenario could be examined through its search.   
However, domain wall problem makes models invalid in many cases \cite{dw}. 
It is caused by the spontaneous breaking of a discrete symmetry which 
remains as an exact one through the explicit breaking of $U(1)_{PQ}$ 
due to QCD anomaly. 

A well-known simple model without the domain wall problem is the KSVZ 
model \cite{ksvz}.
This model has a pair of extra color triplet fermions $(D_L, D_R)$ with
the PQ charge and a Yukawa coupling $S\bar D_LD_R$ with a singlet scalar $S$. 
Although domain walls bounded by the string due 
to the $U(1)_{PQ}$ breaking are generated in this model, 
the domain wall problem is not caused because they are not 
topologically stable \cite{stdw}.
Moreover, after the spontaneous breaking of $U(1)_{PQ}$,
the model could have an effective discrete symmetry which could be violated 
only through the QCD anomaly depending on the charge assignment 
to the fields.
This effective symmetry could play an interesting role 
in low energy phenomena.\footnote{There are several articles which study a 
phenomenological role of the remnant discrete symmetry of $U(1)_{PQ}$ 
\cite{pqscot}.}
In this paper, we consider this kind of possibility for the PQ symmetry
without introducing extra colored fermions. 
We require the model to be free from the domain wall problem and 
have the above mentioned remnant effective symmetry which could be responsible 
in the low energy phenomena.
   
We start our discussion with examining the DFSZ model \cite{dfsz} 
as a typical example.
It has a singlet scalar $S$ and two doublet Higgs scalars $H_u$ and $H_d$. 
These doublet scalars have weak hypercharge with reverse sign so as to couple 
with the up- and down- quarks, respectively. 
Yukawa couplings and scalar potential in this model are fixed as \cite{dfsz}
\begin{eqnarray}
-{\cal L}_{PQ}^y&=&\sum_{i,j=1}^3\left(y^u_{ij}H_u\bar q_{L_i}u_{R_j}
+y^d_{ij}H_d\bar q_{L_i}d_{R_j}
+y^e_{ij}H_d\bar \ell_{L_i}e_{R_j} +{\rm h.c.}\right), \nonumber \\
V_{PQ}&\supset&\frac{\kappa}{2}\left(S^2[H_uH_d]+{\rm h.c.}\right)+\cdots,
\label{dfs}
\end{eqnarray}
where $[\psi\phi]$ in $V_{PQ}$ stands for the abbreviation of 
$\epsilon_{ij}\psi_i\phi_j$.
This model could have a global $U(1)_{PQ}$ symmetry which is broken through
the QCD anomaly. If we represent its charge of the field $f$ as $X_f$,
the $U(1)_{PQ}$ invariance of quark Yukawa couplings in ${\cal L}_{PQ}^y$ 
requires to satisfy
\begin{equation}
H_u-X_{q_i}+X_{u_j}=0, \qquad H_d-X_{q_i}+X_{d_j}=0.
\label{condpq}
\end{equation}
On the other hand, if the quarks are transformed for this $U(1)_{PQ}$ as 
\begin{equation}
q_{L_i}\rightarrow e^{iX_{q_i}\alpha}q_{L_i},\quad 
u_{R_i}\rightarrow e^{iX_{u_i}\alpha}u_{R_i}, \quad
d_{R_i}\rightarrow e^{iX_{d_i}\alpha}d_{R_i},
\end{equation}
the QCD parameter $\theta_{\rm QCD}$ is shifted as a result of anomaly as
\begin{equation}
\theta_{\rm QCD} \rightarrow \theta_{\rm QCD}
+\frac{1}{2}\sum_{i=1}^3\left(X_{u_i}+X_{d_i}-2X_{q_i}\right)\alpha
=\theta_{QCD} - \frac{3}{2}(X_{H_u}+X_{H_d})\alpha, 
\label{theta}  
\end{equation}
where eq.~(\ref{condpq}) is used.
Since this $U(1)_{PQ}$ is assumed to have the QCD anomaly,
$X_{H_u}+X_{H_d}\not=0$ should be satisfied.
In this context, a term explicitly shown in $V_{PQ}$ is essential 
for this model since it requires that the $S$ should have 
$U(1)_{PQ}$ charge $2X_S=-X_{H_u}-X_{H_d}$.
As a result, the $U(1)_{PQ}$ is spontaneously 
broken through the VEV of $S$, and 
the pseudo-Nambu-Goldstone boson (axion) associated to this breaking 
could solve the strong $CP$ problem \cite{axion}.
If the axion decay constant $f_a$ satisfies 
$10^9~{\rm GeV}~{^<_\sim}~f_a~{^<_\sim}~10^{12}~{\rm GeV}$,
any cosmological and astrophysical problems are known not to be 
caused except for the domain wall problem \cite{fa}.

Since $\theta_{\rm QCD}$ has a period $2\pi$ in eq.~(\ref{theta}), 
$\alpha$ should be written as 
$\alpha=2\pi\frac{k}{N}$ where $k=1,\cdots,N(\equiv 3X_S)$ for $X_S$, which 
is normalized to be an integer. 
The model is found to have a $Z_N$ symmetry, which is a subgroup of $U(1)_{PQ}$
and corresponds to the symmetry among $N$ degenerate 
vacua \cite{dw}.\footnote{The axion decay constant $f_a$ is connected to the
PQ symmetry breaking scale $\langle S\rangle$ by $f_a={\langle S\rangle}{N}$.} 
Since topologically stable domain walls are generated among these vacua
for $N>1$ and they overclose the universe, 
such models are ruled out cosmologically. 
In order to escape this situation, one may consider the model with $N=1$ or 
the introduction of a suitable explicit breaking of $Z_N$ which 
resolves the degeneracy among vacua for $N>1$ \cite{dw,explicit}. 

Here we consider a PQ charge assignment which realizes
both $N=1$ and the existence of an effective $Z_2$ symmetry starting from
the DFSZ model.
For that purpose, we may consider a possibility that only a part of quarks 
has the PQ charge by introducing a Higgs doublet with no PQ charge. 
In such a case, we may find $N=X_S$ from the above discussion.
On the other hand, the existence of the effective $Z_2$ symmetry requires
$|X_S|=2$ for the PQ charge normalized as an integer.
This suggests that it seems to be difficult to solve the domain wall 
problem imposing the existence of an effective remnant symmetry $Z_2$
without introducing extra colored fermions as long as the doublet 
Higgs is assumed to have the PQ charge. 

From this view point, in the following part, we consider 
the PQ charge assignment in the SM framework supplemented 
only by a singlet scalar $S$. 
First, we assign the PQ charge to the quarks so as not to cause the
domain wall problem, that is, to realize $N=1$ where $N$ is calculated as
$N=\frac{1}{2}\sum_f(X_{f_R}-X_{f_L})$. 
Using this PQ charge assignment, we study quark mass matrices. 
Next, in order to extend the model to the lepton sector,
we introduce an extra doublet scalar $\eta$ and 
three right-handed neutrinos $N_i$.  
After the PQ symmetry breaking at a high energy scale such 
as $10^9{\rm GeV}~{^<_\sim}~\langle S\rangle~{^<_\sim}~10^{12}{\rm GeV}$, 
the effective model is shown to be reduced to the scotogenic 
neutrino mass model proposed by Ma \cite{ma}. 
The resulting model has no strong $CP$ problem, and has favorable quark 
mass hierarchy and CKM mixing in addition to several features of the 
original scotogenic model such as the explanation of neutrino mass generation, 
leptogenesis and the existence of dark matter \cite{radnm,tribi,stabf1,ks}. 

\begin{figure}[t]
\begin{center} 
\begin{tabular}{c||cccccccccc}\hline
&$q_{L_1}$ & $q_{L_2}$ & $q_{L_3}$ &$u_{R_1}$& $u_{R_2}$ &$u_{R_3}$ & 
 $d_{R_1}$ & $d_{R_2}$ & $d_{R_3}$ & $S$  \\ \hline 
$X_f$ & $-4$ & $-2$ & 0 & 4 & 2 & 0 & $-10$ & $-8$ & 2 & $-2$  \\ 
$Z_2$ & $+$ & $+$ & $+$ &$+$ & $+$ & $+$ &$+$ & $+$ & $+$ & 
$+$  \\ \hline
\end{tabular}
\end{center}

{\footnotesize {\bf Table~1}~~ A summary of $U(1)_{\rm PQ}$ charge of 
quarks. Other field contents of the SM have no PQ charges. 
Parity for the remnant effective $Z_2$ is also shown.}  
\end{figure}

We modify the DSFZ model without introducing new colored fermions 
but changing the PQ charge assignment to the quarks as shown 
in Table~1. As in the SM, the model has only one Higgs doublet $\phi$
which has no PQ charge. The singlet scalar $S$ has the PQ charge.
Since $\phi$ has no PQ charge, quark Yukawa couplings are 
allowed only for the top quark as a renormalizable term. 
However, if we take account of 
nonrenormalizable Yukawa couplings containing the singlet scalar $S$,
the following couplings are found to be allowed,
\begin{eqnarray}
-{\cal L}_y^q&=&\sum_{i=1}^3\left[\sum_{j=1}^3y^u_{ij}\left(\frac{S}{M_\ast}
\right)^{\frac{1}{2}(X_{u_{R_j}}-X_{q_{L_i}})}\bar q_{L_i}\phi u_{R_j}
+\sum_{j=1}^2y^d_{ij}\left(\frac{S^\ast}{M_\ast}\right)^{\frac{1}{2}(X_{d_{R_j}}-X_{q_{L_i}})}
\bar q_{L_i}\tilde\phi d_{R_j} \right. \nonumber \\
&+&\left. y^d_{i3}\left(\frac{S}{M_\ast}\right)^{\frac{1}{2}(X_{d_{R_3}}-X_{q_{L_i}})}
\bar q_{L_i}\tilde\phi d_{R_3} + {\rm h.c.}\right], 
\label{yukawa0}
\end{eqnarray} 
where $\tilde\phi=i\tau_2\phi^\ast$ and $M_\ast$ is a cut-off scale 
of the model.
In this case, eq.~(\ref{theta}) is easily found to be written as
\begin{equation}
\theta_{\rm QCD} \rightarrow \theta_{QCD} + \alpha.
\end{equation}
This shows that the model has no degenerate vacua ($N=1$) and then  
no domain wall problem exists.

After the PQ symmetry breaking, the above nonrenormalizable Yukawa terms
induce a suppression factor to each Yukawa coupling 
which is determined by the PQ charge just like
Froggatt-Nielsen mechanism \cite{fn}.\footnote{The possibility to identify
the PQ symmetry with the Froggatt-Nielsen symmetry has already been 
discussed in \cite{fn-pq}.}
If we take account of the PQ charge in Table~1, 
the quark mass matrices defined by $\bar u_L{\cal M}_uu_R$ and 
$\bar d_L{\cal M}_dd_R$ can be written as
\begin{equation}
{\cal M}_u=\left(\begin{array}{ccc}
y_{11}^u~\epsilon^4 & y_{12}^u~\epsilon^3 & y_{13}^u~\epsilon^2 \\
y_{21}^u~\epsilon^3 & y_{22}^u~\epsilon^2 & y_{23}^u~\epsilon \\
y_{31}^u~\epsilon^2 & y_{32}^u~\epsilon & y_{33}^u \\
\end{array}\right)\langle \phi\rangle , \quad
{\cal M}_d=\left(\begin{array}{ccc}
y_{11}^d~\epsilon^3 & y_{12}^d~\epsilon^2 & y_{13}^d~\epsilon^3 \\
y_{21}^d~\epsilon^4 & y_{22}^d~\epsilon^3 & y_{23}^d~\epsilon^2 \\
y_{31}^d~\epsilon^5 & y_{32}^d~\epsilon^4 & y_{33}^d~\epsilon \\
\end{array}\right)\langle \phi\rangle ,
\end{equation}
where $\epsilon\equiv\frac{|\langle S\rangle|}{M_\ast}$ is used.
In order to examine the features of these mass matrices, 
we take an assumption for simplification such as 
\begin{eqnarray}
&&y_{11}^u=y_{23}^u=y_{32}^u=y_{33}^u=1,\quad
y_{13}^u=y_{22}^u=y_{31}^u=0.1, \quad
y_{12}^u=y_{21}^u=0.7,   \nonumber  \\
&&y_{21}^d=y_{22}^d=y_{31}^d=y_{32}^d=1, \quad y_{11}^d=y_{13}^d=y_{23}^d=0.1,
\quad y_{12}^d=0.022, \quad  y_{33}^d=0.3.
\label{yukawac}
\end{eqnarray}
If we assume $\epsilon=0.08$ in these matrices, 
we can obtain the quark mass eigenvalues and the CKM matrix as follows,
\begin{eqnarray}
&&m_u=2.6~{\rm MeV}, \quad m_c=1.1~{\rm GeV}, \quad 
m_t=174~{\rm GeV}, \nonumber \\ 
&&m_d=6.7~{\rm MeV}, \quad m_s=92~{\rm MeV}, \quad 
m_b=4.2~{\rm GeV}, 
\label{qmass}
\end{eqnarray} 
and
\begin{equation}
V_{\rm CKM}=\left(\begin{array}{ccc}
0.974 &-0.226 & -0.00515 \\
0.226 & 0.974 & -0.0184 \\
0.00918 & 0.0168 & 0.9998 \\
\end{array}\right).
\label{ckm}
\end{equation}
Although the number of free parameters are restricted to 8 in the present case
and we do not make severe tuning for them as shown in eq.~(\ref{yukawac}),
these results seem to be rather good \cite{pdg}.

Here we order some remarks related to the present PQ charge assignment. 
Since the PQ charge is assigned to the quarks in the flavor dependent way,
the axion-quarks couplings are not diagonal to induce the flavor changing
neutral current processes such as $K^\pm\rightarrow \pi^\pm a$ 
and then the PQ breaking scale $\langle S\rangle$
is constrained by them. This problem is studied in \cite{fn-pq}.
If we follow their analysis, the PQ breaking scale $\langle S\rangle$ is 
found to have to satisfy $\langle S\rangle~{^>_\sim}~8\times 10^{10}$~GeV
in the present scenario.
It gives a stronger constraint on $\langle S\rangle$ than the 
astrophysical one.
Apart from the above interesting feature, it is a crucial problem 
how this axion model could be distinguished from others based 
on different PQ charge assignments. 
On this point, we should note that the axion nature determined by the
PQ charge assignment might be experimentally examined through 
the axion-photon coupling \cite{agg,lmn}.
The present model predicts it as 
\begin{equation}
g_{a\gamma\gamma}=\frac{m_a}{\rm eV}
\frac{2.0}{10^{10}{\rm GeV}}\times 1.75.
\end{equation}
It is also useful to note that this relation is not affected 
if the fourth generation quarks exist in this model. 
This is because the domain wall free requirement
imposes their PQ charge should be vector-like.
 
In order to apply this model to the lepton sector,
we combine it with the scotogenic neutrino mass model.
For this purpose, we introduce a new doublet scalar $\eta$ and 
three right-handed neutrinos $N_i$ and assign the PQ charge them 
as $-1$ and $1$, respectively.
The leptons in the SM are assumed to have no PQ charge.
This charge assignment shows that only the $\eta$ and $N_i$ have 
odd parity of the $Z_2$ symmetry which remains as an effective symmetry 
of the model. Although it is broken by the QCD anomaly, 
this breaking does not affect the lepton sector. 
Thus, it could play the same role as the $Z_2$ in the scotogenic model.

Yukawa couplings in the lepton sector and scalar potential 
are represented as \cite{ds} 
\begin{eqnarray}
-{\cal L}_y^\ell&=&\sum_{\alpha=e,\mu,\tau}
\sum_{i=1}^3h_{\alpha i}\bar \ell_\alpha\eta N_i + 
\sum_{i=1}^3y_iS\bar N_i^cN_i +{\rm h.c.}, \nonumber \\
V&=&m_S^2S^\dagger S+\kappa_1(S^\dagger S)^2+\kappa_2(S^\dagger S)(\phi^\dagger\phi)
+\kappa_3(S^\dagger S)(\eta^\dagger\eta) \nonumber \\
&+&m_\phi^2\phi^\dagger\phi+m_\eta^2\eta^\dagger\eta 
+\lambda_1(\phi^\dagger\phi)^2
+\lambda_2(\eta^\dagger\eta)^2 
+\lambda_3(\phi^\dagger\phi)(\eta^\dagger\eta) 
+\lambda_4(\phi^\dagger\eta)(\eta^\dagger\phi) \nonumber \\ 
&+&\frac{\lambda_5}{2}\left[\frac{S}{M_\ast}(\eta^\dagger\phi)^2
+{\rm h.c.}\right],
\label{smodel}
\end{eqnarray}
where we add a dimension 5 term to V as the lowest order one.
We find that eqs.~(\ref{yukawa0}) and (\ref{smodel}) present the most 
general Yukawa couplings and scalar potential which contains the
lowest order terms invariant under the assumed symmetry.

After the symmetry breaking due to $\langle S\rangle\not=0$, $N_i$
and $S$ are found to get masses such as 
$M_i=y_i\langle S\rangle$ and $M_S^2=4\kappa_1\langle S\rangle^2$, respectively.
The effective model at the scale below $M_S$ could be obtained 
by integrating out $S$.
If we do it by using the equation of motion for $S$,
we can obtain the corresponding low energy effective model.
Its scalar potential composed of the light scalars can be written as \cite{ds}
\begin{eqnarray}
V_{\rm eff}&=&\tilde m_\phi^2(\phi^\dagger\phi)+\tilde m_\eta^2(\eta^\dagger\eta)
+\tilde\lambda_1(\phi^\dagger\phi)^2
+\tilde\lambda_2(\eta^\dagger\eta)^2
+\tilde\lambda_3(\phi^\dagger\phi)(\eta^\dagger\eta) 
+\lambda_4(\phi^\dagger\eta)(\eta^\dagger\phi) \nonumber \\
&+&\frac{\tilde\lambda_5}{2}\left[(\eta^\dagger\phi)^2 +{\rm h.c.}\right],
\label{effpot}
\end{eqnarray}
where we use the shifted parameters which are defined as 
\begin{eqnarray}
&&\tilde\lambda_1=\lambda_1-\frac{\kappa_2^2}{4\kappa_1}, \qquad
\tilde\lambda_2=\lambda_2-\frac{\kappa_3^2}{4\kappa_1}, \qquad
\tilde\lambda_3=\lambda_3-\frac{\kappa_2\kappa_3}{2\kappa_1}, \nonumber\\
&&\tilde\lambda_5=\lambda_5\frac{\langle S\rangle}{M_\ast}, \qquad
\tilde m_\phi^2=m_\phi^2+\kappa_2\langle S\rangle^2, \qquad 
\tilde m_\eta^2=m_\eta^2+\kappa_3\langle S\rangle^2.
\label{gcoupl}
\end{eqnarray}

This effective model obtained after the spontaneous 
breaking of $U(1)_{PQ}$ is just the original scotogenic model 
with a $Z_2$ symmetry \cite{ma}, which connects the neutrino mass 
generation with the DM existence. 
In the present case, the right-handed neutrinos do not have their masses
in a TeV region but they are considered to be much heavier.
The coupling $\tilde\lambda_5$ which is crucial for 
the one-loop neutrino mass generation is derived from a nonrenormalizable 
term as a result of the $U(1)_{PQ}$ breaking. 
The model also contains the inert doublet scalar $\eta$ whose lightest 
component can be DM since it has odd parity of the remnant $Z_2$.  
It has charged components $\eta^\pm$ and two neutral components $\eta_{R,I}$.
Their mass eigenvalues can be expressed as
\begin{equation}
M_{\eta^\pm}^2=\tilde m_\eta^2 +\tilde\lambda_3\langle\phi\rangle^2, \qquad
M_{\eta_{R,I}}^2=\tilde m_\eta^2+\left(\tilde\lambda_3+\lambda_4
\pm\tilde\lambda_5\right)\langle\phi\rangle^2.
\label{mscalar}
\end{equation}
We suppose $\tilde m_\eta=O(1)$~TeV although it requires fine tuning 
because of $|\langle S\rangle|\gg |\langle\phi\rangle|$. 

\input epsf
\begin{figure}[t]
\begin{center}
\epsfxsize=7.5cm
\leavevmode
\epsfbox{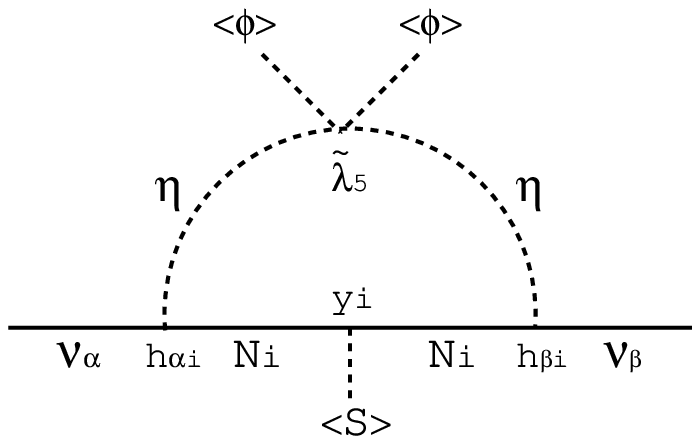}
\end{center}

{\footnotesize {\bf Fig.~1}~~An effective one-loop diagram which generates
neutrino masses through the PQ symmetry breaking. }
\end{figure}

Phenomenology and the related analysis in the lepton sector is almost 
the same as the one given in \cite{ds} where the extension of the KSVZ 
model is studied. 
However, in order to make the paper self-contained, we briefly review 
important points on neutrino mass generation, leptogenesis 
and DM relic abundance and also add some analyses which are changed 
from the ones given in \cite{ds} here. 
\vspace*{5mm}

\noindent
\underline{Neutrino mass generation}

The model contains the heavy right-handed neutrinos, whose Yukawa couplings 
with the doublet leptons $\ell_\alpha$ and the inert doublet scalar $\eta$ 
are shown in the above ${\cal L}_y^\ell$.
However, neutrino masses are not generated at tree-level 
since $\eta$ is assumed to have no VEV. 
They can be generated radiatively through the one-loop diagram shown in Fig.~1 
since both the right-handed neutrino masses and the mass difference 
between $\eta_R$ and $\eta_I$ are induced after 
the $U(1)_{PQ}$ breaking. 
The latter is generated through the 
$\tilde\lambda_5(\eta^\dagger\phi)^2 $ as easily found from eq.~(\ref{mscalar}). 
If we note that $M_{\eta_{R,I}}^2\gg |M_{\eta_R}^2-M_{\eta_I}^2|$ is satisfied in
the present model, the neutrino mass formula can be approximately 
expressed as
\begin{equation}
{\cal M}_{\alpha\beta}=\sum_i h_{\alpha i}h_{\beta i}\Lambda_i, \qquad
\Lambda_i\simeq \frac{\tilde\lambda_5\langle\phi\rangle^2}{8\pi^2M_i}
\ln\frac{M_i^2}{\bar M_\eta^2},
\label{lnmass}
\end{equation} 
where $\bar M_\eta^2= \tilde m_\eta^2
+\left(\tilde\lambda_3+\lambda_4\right)\langle\phi\rangle^2$. 

For simplicity, we assume the flavor structure of
neutrino Yukawa couplings $h_{\alpha i}$ which induces the 
tri-bimaximal mixing.  It is considered to be rather good 
approximation for the lepton mixing for the purpose of the present study. 
It can be realized for \cite{tribi}
\begin{equation}
h_{ej}=0, \quad h_{\mu j}=h_{\tau j}\equiv h_j \quad (j=1,2); \qquad 
h_{e3}=h_{\mu 3}=-h_{\tau 3}\equiv h_3, 
\label{flavor}
\end{equation}
where the charged lepton mass matrix is assumed to be diagonal.
In that case, the mass eigenvalues are estimated as 
\begin{eqnarray}
&&m_1=0, \qquad m_2= 3|h_3|^2\Lambda_3, \nonumber \\
&&m_3=2\left[|h_1|^4\Lambda_1^2+|h_2|^4\Lambda_2^2+
2|h_1|^2|h_2|^2\Lambda_1\Lambda_2\cos 2(\theta_1-\theta_2)
\right]^{1/2}, 
\label{nmass}
\end{eqnarray}
where $\theta_j={\rm arg}(h_j)$. We find that the squared mass differences 
required by the neutrino oscillation data could be derived \cite{pdg} 
if the parameters relevant to the neutrino masses are fixed as
\begin{eqnarray}
&&M_1= 10^8~{\rm GeV}, \qquad M_2=4\times 10^8~{\rm GeV}, 
\qquad M_3= 10^9{\rm GeV},  \nonumber \\
&&|h_1|= 10^{-4.5},  
\qquad |h_2|\simeq 7.2\times 10^{-4}\tilde\lambda_5^{-0.5}, \qquad 
|h_3|\simeq 3.1\times 10^{-4}\tilde\lambda_5^{-0.5}, 
\label{yukawa}
\end{eqnarray}
for $\tilde m_\eta=1$~TeV, for example.
\vspace*{5mm}

\noindent
\underline{Leptogenesis}

The mass formula (\ref{nmass}) and the condition (\ref{yukawa}) suggest 
that the requirement for neutrino masses could be satisfied by using 
two right-handed neutrinos only.
Thus, both mass and neutrino Yukawa couplings of a remaining 
right-handed neutrino are free from the neutrino oscillation data
if its contribution to the neutrino mass is negligible.
Such a situation is found to be realized for 
$|h_1|^2\Lambda_1 \ll |h_2|^2\Lambda_2$ in eq.~(\ref{nmass}).
It is suitable for thermal leptogenesis \cite{leptg} since a 
sufficiently small neutrino Yukawa coupling $h_1$ makes 
the out-of-equilibrium decay of $N_1$ efficient.

Using the values of parameters shown in eq.~(\ref{yukawa}), 
we find the required baryon number asymmetry could be generated 
for $M_1~{^>_\sim}~10^8$~GeV through the decay of the thermal $N_1$ 
by solving the Boltzmann equation. 
An interesting point is that the lightest right-handed neutrino mass 
could be smaller than the Davidson-Ibarra bound \cite{di}
in the ordinary thermal leptogenesis.
We have $Y_B\left(\equiv\frac{n_B}{s}\right)
>8.3\times 10^{-11}$ for a maximal $CP$ phase 
in the $CP$ violation parameter $\varepsilon$ and
the parameter set given in (\ref{yukawa}) 
if $\tilde\lambda_5$ takes a value in the range 
$8\times 10^{-4}~{^<_\sim}~\tilde\lambda_5~{^<_\sim}~2\times 10^{-2}$ \cite{ds}. 
An assumed value $\epsilon=0.08$ in eqs.~(\ref{qmass}) 
and (\ref{ckm}) requires $\lambda_5$ to be in the range 
$0.01~{^<_\sim}~\lambda_5~{^<_\sim}~0.25$. 
It seems to be a natural value 
as a coefficient of the nonrenormalizable term and also to be 
a consistent value in comparison with the values used in eq.~(\ref{yukawac}).
\vspace*{5mm}

\noindent
\underline{DM relic abundance}

As is well known, the axion can be a good DM candidate. Its energy density 
in the present universe is estimated as\cite{explicit}
\begin{equation}
\Omega_ah^2=2\times 10^4\left(\frac{\langle S\rangle}
{10^{16}~{\rm GeV}}\right)^{7/6}\langle\theta_i^2\rangle,
\end{equation}
where $\langle\theta_i\rangle$ stands for the initial axion misalignment.
The axion contribution to the DM abundance crucially depends on 
$\langle\theta_i\rangle$. 
We find that it could be too small to give the required value 
$\Omega_{\rm DM} h^2=0.12$ \cite{pdg} 
for $\langle S\rangle< 10^{11}$~GeV
even if we assume $\langle\theta_i\rangle=O(1)$.
However, the model has a unique $Z_2$ odd field $\eta$ among the weak 
scale fields. The lightest neutral component $\eta_R$ could be a good 
DM candidate in case of $\lambda_4, \tilde\lambda_5<0$. 
In fact, since $\tilde m_\eta$ is assumed to be of $O(1)$~TeV in this model, 
the mass of each component of $\eta$ could be degenerate sufficiently 
for wide range values of $\tilde\lambda_3$ and $\lambda_4$ as 
found from eq.~(\ref{mscalar}). 
This makes the coannihilation among the components of $\eta$ 
effective enough to reduce the $\eta_R$ abundance \cite{inert2,ks}.
Thus, if the couplings $\tilde\lambda_3$ and $\lambda_4$ take suitable values,  
the relic abundance of $\eta_R$ can be tuned to the required value.
In the left panel of Fig.~2, such points in the 
$(\tilde\lambda_3,~\lambda_4)$ plane are plotted by a red solid line. 
The figure shows that $\tilde\lambda_3$ and/or $|\lambda_4|$ are 
required to take rather large values for realization of 
the DM abundance.\footnote{The axion could be the dominant component 
of DM for $\langle S\rangle>10^{11}$~GeV and $\langle\theta_i\rangle=O(1)$. 
In that case, we need to consider larger $|\lambda_{3,4}|$ to reduce the
$\eta_R$ abundance. In this paper, we focus our study on the $\eta_R$ 
dominated DM case.} 
  
A problem related to these values is their influence to the vacuum 
stability and the perturbativity of the model.
It can be examined by solving the renormarization group equations 
(RGEs) for the quartic scalar couplings. 
Since the model has no extra colored fermion and the running of 
gauge couplings changes from the one found in \cite{ds}, 
we should take account of it
in the present analysis. Here we note that vacuum stability conditions 
in this model are given at a scale below $M_S$ as \cite{pstab}
\begin{equation}
\tilde\lambda_1>0, \qquad \tilde\lambda_2>0, \qquad
\tilde\lambda_3>-2\sqrt{\tilde\lambda_1\tilde\lambda_2},  \qquad
\tilde\lambda_3+\lambda_4-|\tilde\lambda_5|>
-2\sqrt{\tilde\lambda_1\tilde\lambda_2}.
\label{instab}
\end{equation}
At a scale above $M_S$, in addition to the same conditions 
for $\lambda_{1,2,3}$ as eq.~(\ref{instab}) except for the last one, 
new conditions    
\begin{equation}
\kappa_1>0, \qquad
\kappa_2>-2\sqrt{\lambda_1\kappa_1}, \qquad
\kappa_3>-2\sqrt{\lambda_2\kappa_1},
\label{stability2}
\end{equation}
should be satisfied. The couplings in both regions are connected through
eq.~(\ref{gcoupl}). 

If we use $\tilde\lambda_1$ determined by 
the observed Higgs mass, we can find an a region in 
the $(\tilde\lambda_3,~\lambda_4)$ plane allowed by the stability 
for a fixed $\tilde\lambda_2$ by applying the last condition 
in eq.~(\ref{instab}).
Here we should remind that $\tilde\lambda_5$ is restricted to small values
through the leptogenesis.
In the left panel of Fig.~2, we plotted lines
$\tilde\lambda_3+\lambda_4-|\tilde\lambda_5|=
-2\sqrt{\tilde\lambda_1\tilde\lambda_2}$ for $\tilde\lambda_2=0.01$ 
and $0.4$. Points above the line corresponding to each $\tilde\lambda_2$ 
satisfy the stability condition at the weak scale. 
From this figure, we can find values of $\tilde\lambda_3$ and $\lambda_4$
which are used as initial values at the weak scale for the RGEs study. 
Large values required for $\tilde\lambda_3$ and/or $|\lambda_4|$ 
are expected to improve the situation for the vacuum stability up to $M_S$ 
compared with the SM since they give positive contributions to 
the $\beta$-function for the quartic coupling $\tilde\lambda_1$. 
On the other hand, their large contributions to the $\beta$-function of 
the scalar quartic couplings $\tilde\lambda_i$ tend to make 
the perturbativity of the model break at a scale below $\langle S\rangle$.
We identify a perturbativity violating scale with the cut-off scale 
$M_\ast$ of the model so that $M_\ast$ is fixed at a scale where 
$\lambda_i(M_\ast)>4\pi$ happens for any $\lambda_i$. 
The validity of the present scenario is guaranteed only for 
$\langle S\rangle<M_\ast$.

\begin{figure}[t]
\begin{center}
\epsfxsize=7.5cm
\leavevmode
\epsfbox{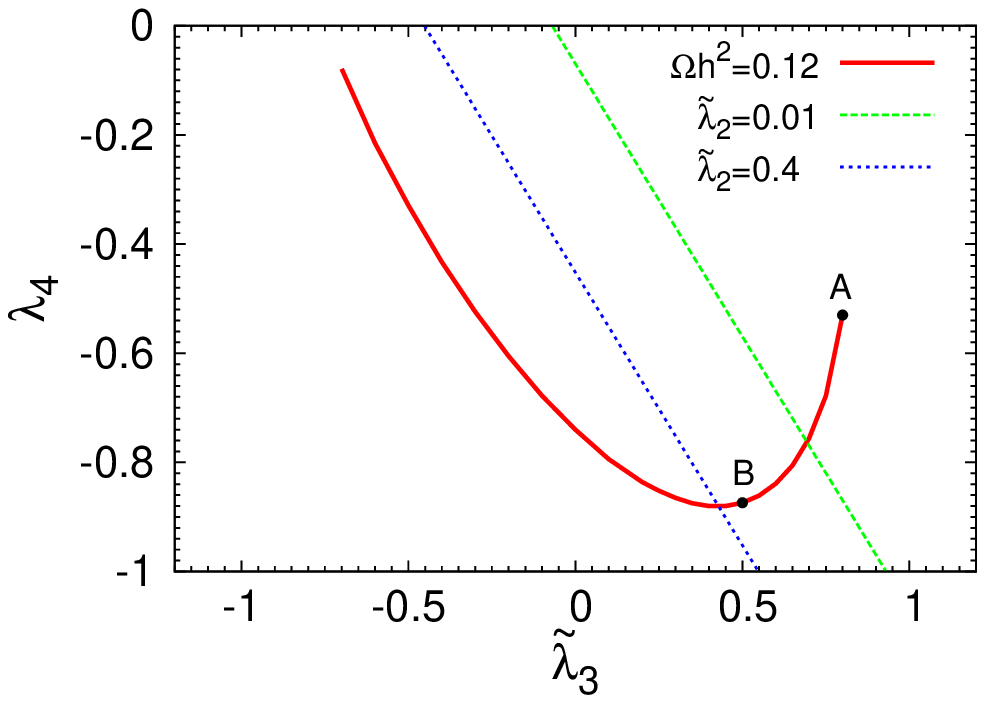}
\epsfxsize=7.5cm
\leavevmode
\epsfbox{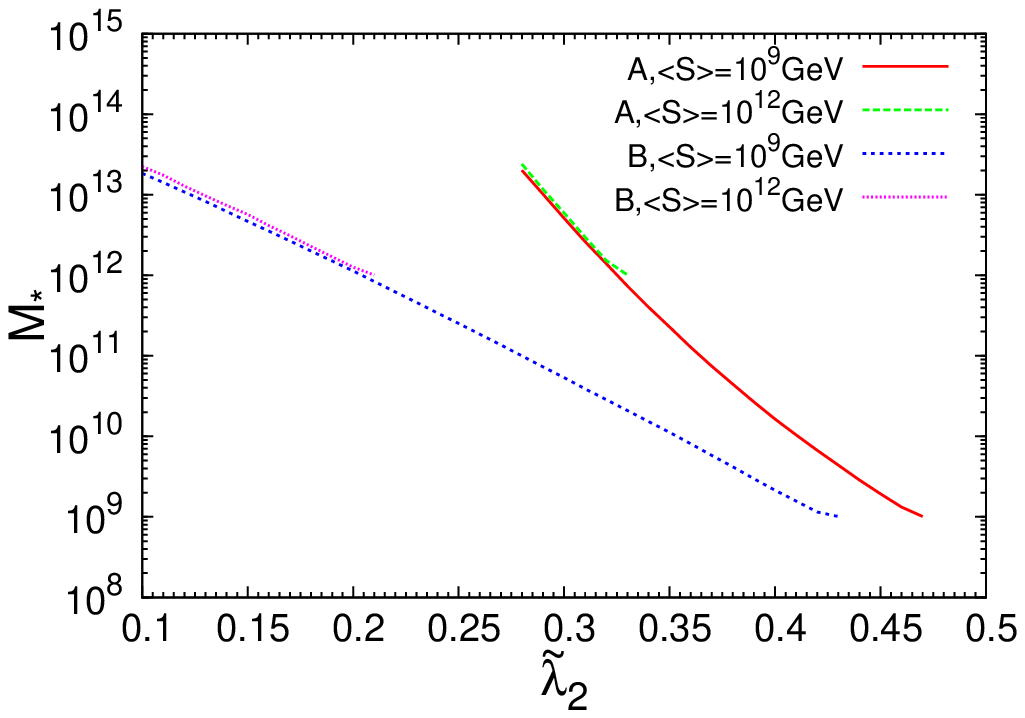}
\end{center}
\vspace*{-3mm}

{\footnotesize {\bf Fig.~2}~~
Left panel: Points plotted by a red solid line in the 
$(\tilde\lambda_3,~\lambda_4)$ plane can realize
the required DM relic abundance $\Omega h^2=0.12$. 
The last condition in eq.(\ref{instab})
is satisfied at a region above a straight line which represents
$\tilde\lambda_3+\lambda_4=|\tilde\lambda_5|-
2\sqrt{\tilde\lambda_1\tilde\lambda_2}$ for a fixed $\tilde\lambda_2$.
Right panel: Cut-off scale $M_\ast$ as a function of $\tilde\lambda_2$
which is fixed as a value at $M_Z$. These are plotted for two points 
A$(0.80, -0.530)$ and B$(0.50,-0.874)$ which are marked by the 
black bulbs in the left panel.}
\end{figure}

We estimate $M_\ast$ by using one-loop RGEs 
and checking the vacuum stability at each scale. 
Since the PQ symmetry breaking scale is constrained through the axion 
physics, $M_\ast$ should be in the range 
$1.25\times 10^{10}{\rm GeV}~{^<_\sim}~M_\ast~{^<_\sim}~
1.25\times 10^{12}{\rm GeV}$ for $\epsilon=0.08$ which is used in this model.
In the right panel of Fig.~2, $M_\ast$ is plotted as a function of 
$\tilde\lambda_2$ for two points A and B in the 
$(\tilde\lambda_3,~\lambda_4)$ plane where the required relic abundance 
is realized. In this study, $M_s=\langle S\rangle$ is assumed and
quartic couplings $\kappa_i$ are fixed as 
$\kappa_1=\frac{M_S^2}{4\langle S\rangle^2}$ and $\kappa_{2,3}=0.1$ 
at $M_S$.\footnote{It is useful to note that larger values of 
$\kappa_{2,3}$ make $M_\ast$ smaller.}
A lower end point in each line corresponds to $M_\ast=\langle S\rangle$.
Upper end points found in lines for the case A stand for the value of 
$\tilde\lambda_2$ where the vacuum stability is violated before reaching 
$M_\ast$. 
This figure shows that $M_\ast$ could 
take a consistent value with $\epsilon$ used here and 
$\langle S\rangle$ in the range imposed by the axion physics
as long as $\tilde\lambda_2$ takes a suitable value.

Finally, we summarize the paper.
We have proposed an invisible axion model free from the domain wall
problem by taking a novel PQ charge assignment to the quarks.
The PQ charge in the quark sector can play a role of $U(1)$ charge
in the Froggatt-Nielsen mechanism so that both the favorable quark
mass eigenvalues and the CKM matrix are obtained.
An extension of the model to the lepton sector can be done by 
introducing an extra inert doublet scalar $\eta$ and 
three right-handed neutrinos $N_i$ as new ingredients.
After the PQ symmetry breaking, the effective low energy 
model can be identified with
the scotogenic model for neutrino masses. It has an effective 
$Z_2$ symmetry as a remnant of the $U(1)_{PQ}$ symmetry, which is violated 
only through the QCD anomaly.
Under this $Z_2$ symmetry, only these new ones $\eta$ and $N_i$ 
have its odd parity and the lightest one could be a good DM candidate 
just as the scotogenic neutrino mass model.
In the similar way as it, the neutrino masses are generated at one-loop 
level and the DM abundance can be explained by the relic neutral 
component of $\eta$. 

The baryon number asymmetry could be generated through
the out-of-equilibrium decay of the lightest right-handed neutrino 
just as the ordinary thermal leptogenesis in the tree-level seesaw model.
However, the mass bound of the lightest right-handed neutrino
could be relaxed. 
This simple extension can relate the strong $CP$ problem to
the quark mass hierarchy and also the origin of neutrino masses and DM. 
The model might suggest a hint for a new way to construct models 
beyond the SM.

\section*{Acknowledgements}
I would like to thank Dr.~Hamaguchi and Dr.~Redigolo for informing their
works \cite{fn-pq}. This work is partially supported by MEXT Grant-in-Aid 
for Scientific Research on Innovative Areas (Grant No. 26104009).

\newpage
\bibliographystyle{unsrt}

\end{document}